\documentclass[sigconf,natbib=false]{acmart}
\usepackage{fancyhdr}
\usepackage{graphicx}
\usepackage{booktabs}
\usepackage{multirow}
\usepackage{enumitem}
\usepackage{bbm}
\usepackage{colortbl}
\usepackage{amsmath}

\newcommand{\best}{\cellcolor[HTML]{C5DFF8}}
\newcommand{\second}{\cellcolor[HTML]{E6FFFD}}
\newcommand{\figref}[1]{Figure~\ref{#1}}

\AtBeginDocument{%
  }

\copyrightyear{2024}
\acmYear{2024}
\setcopyright{acmlicensed}\acmConference[CIKM '24]{Proceedings of the 33rd ACM International Conference on Information and Knowledge Management}{October 21--25, 2024}{Boise, ID, USA}
\acmBooktitle{Proceedings of the 33rd ACM International Conference on Information and Knowledge Management (CIKM '24), October 21--25, 2024, Boise, ID, USA}
\acmDOI{10.1145/3627673.3679919}
\acmISBN{979-8-4007-0436-9/24/10}

\RequirePackage[
  datamodel=acmdatamodel,
  style=acmnumeric,
  ]{biblatex}

\begin{document}

\title{Evolving to the Future: Unseen Event Adaptive Fake News Detection on
Social Media}


\author{Jiajun Zhang}
\email{zhangjiajun519@gmail.com}
\affiliation{%
  \institution{Beijing Institute of Technology}
  \city{Beijing}
  \country{China}
}

\author{Zhixun Li}
\email{zxli@se.cuhk.edu.hk}
\affiliation{%
  \institution{The Chinese University of Hong Kong, China}
  \city{Hong Kong}
  \country{China}
}

\author{Qiang Liu}
\email{qiang.liu@nlpr.ia.ac.cn}
\affiliation{%
  \institution{Institute of Automation, Chinese Academy of Sciences}
  \city{Beijing}
  \country{China}
}

\author{Shu Wu}
\email{shu.wu@nlpr.ia.ac.cn}
\affiliation{%
  \institution{Institute of Automation, Chinese Academy of Sciences}
  \city{Beijing}
  \country{China}
}

\author{Zilei Wang}
\email{zlwang@ustc.edu.cn}
\affiliation{%
  \institution{University of Science and Technology of China}
  \city{Hefei}
  \country{China}
}

\author{Liang Wang}
\email{wangliang@nlpr.ia.ac.cn}
\authornotemark[1]
\affiliation{%
  \institution{Institute of Automation, Chinese Academy of Sciences}
  \city{Beijing}
  \country{China}
}

\renewcommand{\shortauthors}{Jiajun Zhang et al.}
\authornote{Corresponding author}


\begin{abstract}
  With the rapid development of social media, the wide dissemination of fake news on social media is increasingly threatening both individuals and society. One of the unique challenges for fake news detection on social media is how to detect fake news on future events. Recently, numerous fake news detection models that utilize textual information and the propagation structure of posts have been proposed. Unfortunately, most of the existing approaches can hardly handle this challenge since they rely heavily on event-specific features for prediction and cannot generalize to unseen events.
  To address this, we introduce \textbf{F}uture \textbf{AD}aptive \textbf{E}vent-based Fake news Detection (FADE) framework. Specifically, we train a target predictor through an adaptive augmentation strategy and graph contrastive learning to obtain higher-quality features and make more accurate overall predictions. Simultaneously, we independently train an event-only predictor to obtain biased predictions. We further mitigate event bias by subtracting the event-only predictor's output from the target predictor's output to obtain the final prediction. Encouraging results from experiments designed to emulate real-world social media conditions validate the effectiveness of our method in comparison to existing state-of-the-art approaches.
\end{abstract}



\maketitle
\vspace{-15pt}
\section{Introduction}
With the rapid growth of the Internet, social media has become a key platform for sharing opinions and information. However, this has also increased the spread of fake news. In the era of mobile Internet, fake news has become more common and easier to spread. This can shape public opinion, cause economic loss, and lead to serious political consequences \cite{liu2018mining}. Therefore, detecting fake news is a crucial problem that needs to be addressed.

\begin{figure}[t]
\includegraphics[width=0.95\linewidth]{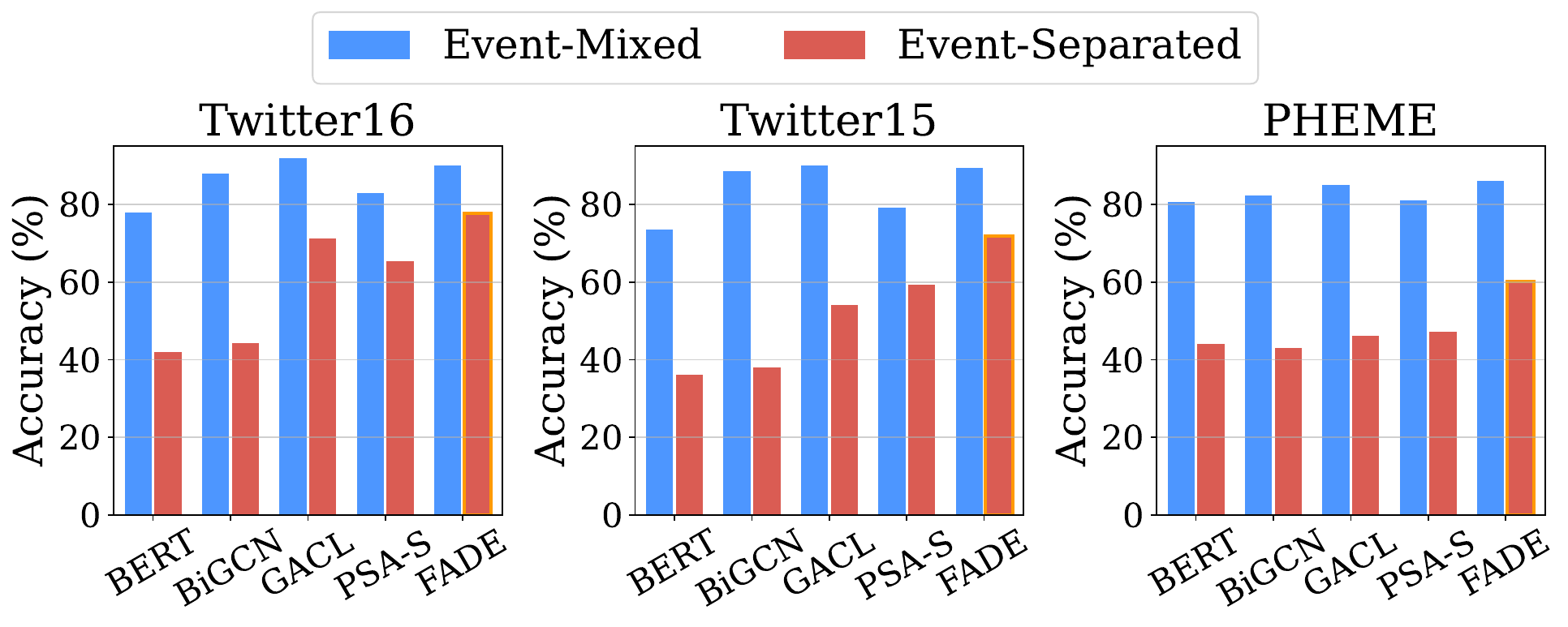}
\caption{Comparison of mean accuracy over 10 runs for each approach in event-mixed and event-separated settings. Note PSA-S is designed for event-separated scenarios.}
\vspace{-25pt}
\label{figure1}
\end{figure}

In real-world social media scenarios, trending events are inherently dynamic and ever-changing. Fake news is often crafted around current hot-button issues that capture public attention. Therefore, an effective fake news detection model should be trained on news reporting past events and be capable of detecting fake news related to future events. In other words, the training and testing data for a detection model are non-independent and identically distributed (non-iid). \textcite{eann} initially addressed generalization for future events in multimedia fake news detection. \textcite{zhu2022generalizing} and \textcite{psa} explored entity and event generalization in text-based fake news detection, respectively. However, most text graph-based methods \cite{rvnn,gcan,bigcn,yang2021rumor,gacl,ma2022towards,recl} overlook generalization for unseen events, opting for event-mixed data splits. This approach significantly overestimates their detection capabilities. In event-separated scenarios \cite{psa}, where models are tested on completely new events, accuracy falls by over 40\%, as depicted in \figref{figure1}. This highlights a major shortfall in current methods' ability to detect fake news from unseen events in real-world settings.

We believe that the failure of existing models in detecting fake news in unseen events can be attributed to two main reasons: (1) \textbf{Insufficient quality of learned features:} Within each event, numerous samples are sharing highly similar keywords, but news from different events often lack these keywords. For example, in the Twitter15 dataset, among all 48 news samples in the event 'E689', 46 share the keywords 'white house' and 'rainbow'. Additionally, news from different events often exhibit vastly different propagation structures. For instance, news about celebrity gossip or popular culture tends to form flat propagation trees, whereas news on political or social issues often results in trees with greater depth. These factors lead models to learn event-specific features rather than shared features across all events, resulting in insufficient quality of the learned features and representations. (2) \textbf{Lack of effective debiasing techniques:} News samples within the same event often have consistent labels. For instance, in the Twitter dataset, large-size events encompass more than 70\% samples, with each event’s samples invariably having the same class label. Combined with the distinct propagation structures and node attributes mentioned in reason 1, this leads to a severe event bias in the dataset. Existing methods \cite{eann, yang2021rumor, lin2022detect, gacl, ma2022towards} have achieved some success using adversarial debiasing and data augmentation techniques, but they still fail to perform effectively in event-separated fake news detection.

To effectively detect fake news in future unseen events, we propose the \textbf{F}uture \textbf{AD}aptive \textbf{E}vent-based Fake news Detection (FADE) framework for fake news detection in this paper. Our framework consists of a target predictor and an event-only predictor, each trained independently. (1) \textbf{Target Predictor:} Data augmentation is a common training strategy that enhances model robustness by generating diverse training samples \cite{simard2003best,singh2018hide}. We propose an efficient graph augmentation strategy named adaptive augmentation, which generates the most challenging augmented samples in the representation space. We then use high-quality augmented training data to train the target predictor through graph contrastive learning, enabling the predictor to learn higher-quality features. (2) \textbf{Event-Only Predictor:} Common debiasing methods like adversarial debiasing and reweighting \cite{schuster2019towards, xu2023counterfactual,chen2023bias}, typically employed during training, are not suitable for fake news detection due to the excessive number of event categories involved. To address this challenge, inspired by the Potential Outcomes Model \cite{sekhon2008neyman}, we propose to train an event-only predictor and use it for debiasing during the inference stage. Specifically, in training the event-only predictor, we incorporate an average pooling layer for samples under the same event, enabling it to generate predictions driven by event biases. We regard the prediction from the target predictor as a combination of unbiased features and biases inherent in the news. Consequently, we obtain the final debiased prediction by subtracting the event-label biased prediction from the target predictor's prediction during the inference stage.

Overall, the main contributions can be summarized as follows:
\begin{itemize}[leftmargin=*]
\item We propose an adaptive augmentation strategy that generates challenging augmentations in the representation space, enhancing performance without manual design for different datasets.
\item We introduce an inference stage debiasing method, combining biased predictions to obtain unbiased inferences, enhancing generalizability for unseen events.
\item We are the first to address fake news detection in an event-separated setting effectively, with empirical results showing our FADE framework outperforms existing state-of-the-art baselines.
\end{itemize}

\begin{figure*} 
\centering
\includegraphics[width=1.8\columnwidth]{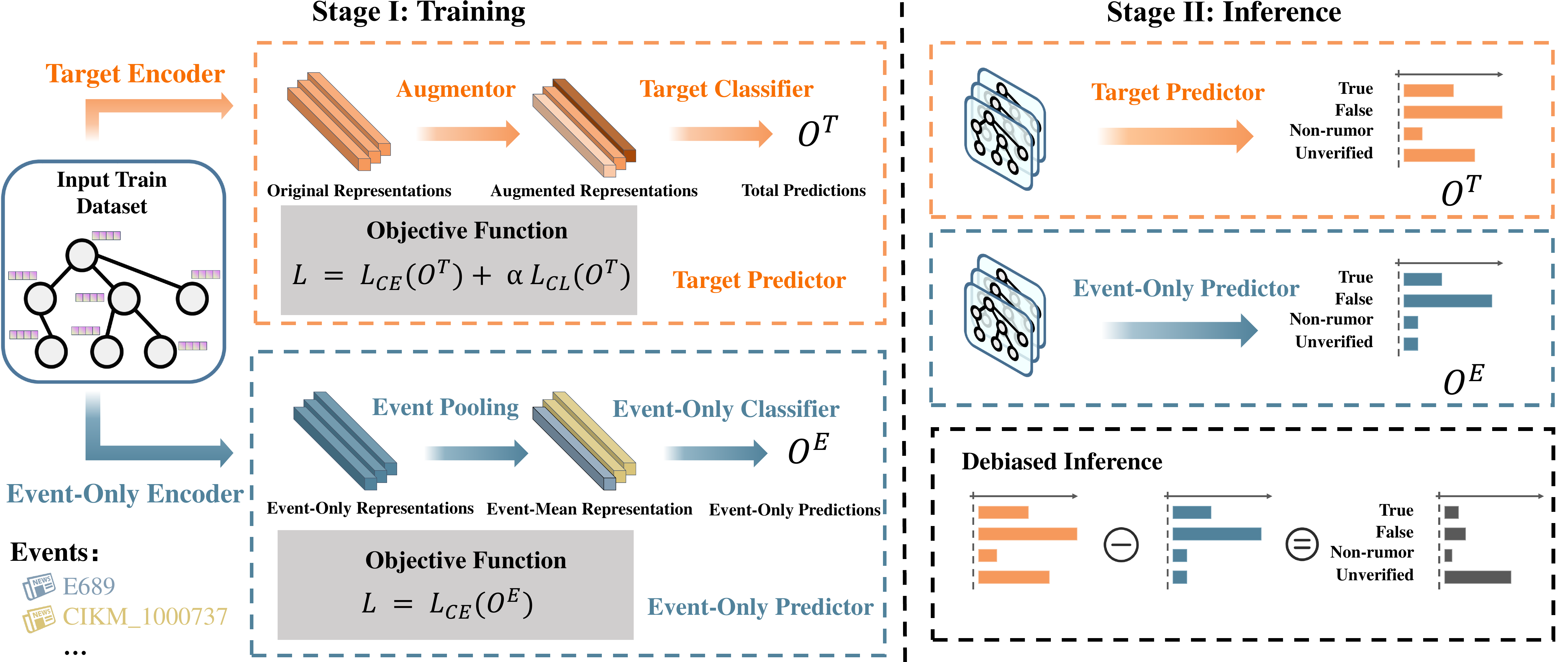}
\caption{Overview of our FADE framework. In the training stage, the target predictor and event-only predictor are trained independently. In the inference stage, we use both predictors' outputs to perform causal debiasing.}
\vspace{-15pt}
\label{figure3}
\end{figure*}

\vspace{-10pt}
\section{PROPOSED APPROACH}
\vspace{-5pt}
\subsection{Problem Definition}
Fake news detection is a classification task aiming to train a classifier on labeled instances and predict labels for unseen instances.

Given a news instance set $\mathcal{C}=\{c_1,c_2,...,c_m \}$ of size $m$, each instance $c_i$ is defined as $c_i=\{r_i,w_1^i,w_2^i,...,w_{n_i-1}^i,P_i \}$. Here, $n_i$ is the number of posts in $c_i$, with $r_i$ as the source post, $w_j^i$ as the $j$-th comment post, and $P_i$ as the propagation structure.

Each instance $c_i$ has a ground-truth label $y_i \in \{R, N\}$ (Rumor or Non-Rumor) and an event label $e_i$. Sometimes, fake news detection is a four-class task with $y_i \in \{N, F, T, R\}$ (Non-rumor, False Rumor, True Rumor, Unverified Rumor). The label $e_i$ represents the event related to $c_i$.

We convert each instance $c_i$ to a graph $G_i = (V_i, \mathbf{X_i}, \mathbf{A_i})$, where $\mathcal{V}_i = \{r_i,w_1^i,w_2^i,...,w_{n_i-1}^i\}$ is the vertex set, $\mathbf{X_i} \in \mathbbm{R}^{n_i \times d}$ are text features embedded by a pre-trained BERT model, and $\mathbf{A_i} \in \{0, 1\}^{n_i \times n_i}$ is the adjacency matrix. Here, $a_{jk}^i = 1$ indicates a reply between post $j$ and post $k$, otherwise $a_{jk}^i = 0$.

$\mathcal{S}=\{(G_1,y_1,e_1), \ldots,(G_m,y_m,e_m)\}$ is the dataset for fake news detection. We define the set of training events as $\mathcal{E}_{tr}$ and test events as $\mathcal{E}_{te}$. If $\mathcal{E}_{tr} \cap \mathcal{E}_{te} \neq \emptyset$, the task is event-mixed fake news detection. If $\mathcal{E}_{tr} \cap \mathcal{E}_{te} = \emptyset$, it is event-separated fake news detection.

\vspace{-10pt}
\subsection{Model Overview}
Figure \ref{figure3} shows the FADE framework, which includes a training stage and an inference stage. In training, the target predictor (GCN-based encoder and classifier) is trained with adaptive augmentation and graph contrastive learning for robust predictions. The event-only predictor is trained with event-mean pooling to capture event bias. In inference, we subtract the event-only predictor's output from the target predictor's to get the final debiased prediction.

\vspace{-10pt}
\subsection{GCN-based Encoder}
We use a Graph Convolutional Network (GCN) \cite{gcn} to extract graph-level representations from structured data. The formula for the $l$-th layer with weight matrix $\mathbf{W}^{(l)}$ is:

\begin{equation} 
\vspace{-5pt}
\label{eq1}
H^{(l+1)} = \sigma\left(\tilde{\mathbf{D}}^{-\frac{1}{2}}\tilde{\mathbf{A}}\tilde{\mathbf{D}}^{\frac{1}{2}}H^{(l)}\mathbf{W}^{(l)}\right),
\end{equation}

where $\tilde{\mathbf{A}} = \mathbf{A} + \mathbf{I}_n$ is the adjacency matrix with self-connections, $\mathbf{I}_N$ is the identity matrix, $\tilde{\mathbf{D}}$ is the degree matrix of $\tilde{\mathbf{A}}$, $\tilde{D}_{ii} = \sum_{j} \tilde{A}_{ij}$, and $H^{0} = X$. $\sigma(\cdot)$ is an activation function. To get graph-level representations from node-level representations, we use:

\begin{equation} 
\vspace{-2pt}
\label{eq2}
R = \text{Pooling}(H^L),
\end{equation}

where $L$ is the number of layers, and the Pooling function is permutation invariant, such as mean or add. $R^O$ denotes the original graph representations. Both the target encoder and the event-only encoder are GCN-based Encoders.

\vspace{-5pt}
\subsection{Adaptive Graph Augmentation}
Data augmentation is a common method to enhance the quality and generalization of model features\cite{krizhevsky2012imagenet,sato2015apac,madnani2010generating}. Currently, most graph augmentation methods \cite{hamilton2017inductive, wang2020graphcrop, you2020graph, rong2019dropedge, zhu2021empirical} rely on random modifications of graph structures or features. However, these methods often fail to optimize model performance and may even accidentally damage significant label-related information \cite{luo2022automated,yue2022label}, leading to decreased classification accuracy. To address this issue, we propose a novel strategy called adaptive augmentation. Specifically, we perform the augmentation in the representation space by adding a perturbation to the original representation $R^O$. In our experiment, we first calculate the centroid and the average Euclidean distance between each original representation and the centroid as $d$ by the following formula:
\begin{equation} \label{eq3}
\vspace{-2pt}
d = \frac{1}{N}\sum\limits_{i=1}^N \Vert \frac{1}{N}\sum\limits_{j=1}^N R_j^O - R_i^O \Vert_2.
\end{equation}
where $N$ denotes the number of samples. Then in the generation process, each time, we stochastically generate multiple random unit vectors. Each unit vector is represented by \(\boldsymbol{\upsilon}\). Then, we use unit vectors to calculate augmented representations for each sample. The augmented representation, denoted as \(R^A\), is computed as:
\begin{equation} \label{eq4}
\vspace{-2pt}
  R^A = R^O + d \boldsymbol{\upsilon}.
\end{equation}

To keep the perturbation intensity reasonable, we use the label \(y\) as a constraint. The target predictor predicts the label \(\hat{y}\) for each augmented news representation. We select the most challenging augmented representation, which is closest to the target classifier's decision boundary, ensuring \(\hat{y} = y\).

\vspace{-5pt}
\subsection{Target Predictor}
In this subsection, we describe the training stage of the target predictor. First, we input $R^O$ into the target classifier for prediction as $O^T = F(R^O)$, where $O^T\in \mathbbm{R}^L$ denotes the predicted class distribution by target classifier ($L$ is the number of class) and $F(\cdot)$ denotes the target classifier. The objective function for the target predictor combines both the contrastive loss and the cross-entropy loss. The cross-entropy loss ($\mathcal{L}_{CE}$) is defined as follow:

\begin{equation} \label{eq6}
\vspace{-3pt}
 \mathcal{L}_{CE} = -\sum\limits_{(R^O_i,y_i)\in \mathcal{S}} CE(\Phi(F(R^O_i),y_i)), 
\end{equation}
where $CE$ denotes cross-entropy loss, $\Phi(\cdot)$ is Softmax. The contrastive loss ($\mathcal{L}_{CL}$) is defined as:

\begin{equation} \label{eq7}
\vspace{-3pt}
\mathcal{L}_{CL} = \frac{-(P^O_i)^T P^A_i}{\Vert P^O_i \Vert_2\Vert P^A_i \Vert_2}.
\end{equation}
Here, we adopt a multi-layer projection head to get projection vectors $P^O$ and $P^A$ from original representations $R^O$ and augmented representations $R^A$. Combining Eq.\ref{eq6}, \ref{eq7}, our overall objective function for the main predictor can be written as follows:
\begin{equation} \label{eq8}
\vspace{-4pt}
\mathop{argmin}\limits_\Theta \mathcal{L} = \mathcal{L}_{CE} + \alpha \mathcal{L}_{CL},
\end{equation}
where $\Theta$ denotes the parameters of the target encoder and classifier, and $\alpha$ denotes the trade-off hyper-parameter to balance contrastive loss and classification loss.

\vspace{-5pt}
\subsection{Event-Only Predictor}
In this subsection, we describe the training stage of the event-only predictor. To train an Event-Only model that generates predictions driven by event-specific features, we incorporate an average pooling layer for samples under the same event. We aggregate the origin representation encoded by the event-only encoder of each sample within event $e_i$ as follows:

\begin{equation} \label{eq9}
\vspace{-4pt}
R^E = Mean(\{R^{\prime O}_j\}_{j=1}^{m_i}),
\end{equation}
where $R^{\prime O}$ denotes the original representation encoded by event-only encoder, $Mean$ denotes the average pooling, and $R^E$ denotes the event-average representation for each event.

Subsequently, we use $R^E$ as the representation for each sample, inputting it into the event-only classifier for prediction. This process yields predictions that are entirely derived from the event bias, as $O^E = F^{\prime}(R^E) $, where $O^E\in \mathbbm{R}^L$ denotes the predicted class distribution by the event-only classifier ($L$ is the number of class) and $F^{\prime}(\cdot)$ denotes the event-only classifier. Then we define the loss function for the event-only predictor as follows:

\begin{equation} \label{eq11}
\vspace{-4pt}
    \mathcal{L}_E = -\sum\limits_{(O^E_i,y_i)\in \mathcal{S}}(CE(\Phi(O^E_i,y_i))
\end{equation}
Then, our overall objective function for the event-only predictor can be written as $\mathop{argmin}\limits_\theta \mathcal{L}_{E}$, where $\theta$ denotes the parameters of the event-only encoder and classifier.

\vspace{-5pt}
\subsection{Debias in inference stage}
After the training stage, we have obtained a target predictor capable of making overall predictions $O^T$ using both unbiased and biased features in news pieces, and an event-only predictor that makes predictions $O^E$ merely based on event biases. 

To reduce event-label bias, inspired by the Potential Outcomes Model\cite{sekhon2008neyman}, we subtract $O^E$ from $O^T$ with a bias coefficient $\beta$ and obtain the debiased output $O^D$.
\begin{equation} \label{eq13}
\vspace{-5pt}
    O^D = O^T - \beta O^E
\end{equation}
$O^D$ reduces biased predictions and retains unbiased ones, thereby achieving a debiasing effect.

\begin{table}[]
\vspace{-5pt}
\renewcommand\arraystretch{0.90}
\resizebox{\linewidth}{!}{
\begin{tabular}{lccc}
\toprule
\textit{Statistic} & Twitter15 & Twitter16 & PHEME  \\ \midrule
\#Source tweets      & 1,490               & 818       & 6,425   \\
\#Events             & 298                 & 182       & 9       \\
\#Users              & 276,663             & 173,487    & 48,843  \\
\#Posts              & 331,612             & 204,820    & 197,852 \\
\#Non-rumors         & 374                & 205       & 4,023   \\
\#False rumors       & 370                & 205       & 2,402   \\
\#Unverified rumors  & 374                & 203       & -      \\
\#True rumors        & 372                & 205       & -      \\ \bottomrule
\end{tabular}}
\caption{Statistics of the datasets}
\vspace{-32pt}
\label{Table1}
\end{table}

\vspace{-5pt}
\section{Experiments}
\vspace{-5pt}
\subsection{Experiment Settings}

We tested our model using the Twitter15, Twitter16, and PHEME fake news datasets from Twitter, detailed in Table \ref{Table1}. We constructed post graphs and generated node embeddings with BERT.

Our data splitting respected event separation across all datasets, preventing overlap in training, testing, and validation events. Validation sets received about 10\% of the data, with the rest split 3:1 between training and testing by event IDs. Twitter15 and Twitter16 were split following \textcite{psa} methods, and PHEME using \textcite{gacl}, ensuring all tests involved unseen events.

\subsubsection{Compared Methods}
\

\textbf{BERT} \cite{bert} is a popular pre-trained model that is used for fake news detection.

\textbf{BiGCN} \cite{bigcn} is a GCN-based model that uses the two key features of news propagation and dispersion to capture the global structure of the news tree.

\textbf{GACL} \cite{gacl} is a GCN-based model using adversarial and contrastive learning for fake news detection.

\textbf{PSA} \cite{psa} is a text-based fake news classifier that can learn writing style and truth stance, thus enhancing its classification capability. \textbf{PSA-S} and \textbf{PSA-M} respectively represent the use of sum and mean as pooling functions.

\textbf{FADE} is our proposed framework for fake news detection using adaptive data augmentation and causal debiasing.

\begin{table}[]
\vspace{-5pt}
\renewcommand\arraystretch{0.95}
\resizebox{\linewidth}{!}{
\begin{tabular}{lccccc}
\toprule
\multicolumn{6}{c}{Twitter15}                                                                                            \\ \midrule
\multicolumn{1}{l|}{\multirow{2}{*}{Method}} & \multicolumn{1}{c|}{\multirow{2}{*}{Acc.}} & U     & N     & T     & F     \\ 
\multicolumn{1}{l|}{}                        & \multicolumn{1}{c|}{}                     & F1    & F1    & F1    & F1    \\ \midrule
\multicolumn{1}{l|}{BERT}                    & \multicolumn{1}{c|}{36.02{\small$\pm$4.80}}                & 40.20{\small$\pm$3.00} & 60.14{\small$\pm$3.30} & 10.23{\small$\pm$5.80} & 25.44{\small$\pm$6.50} \\
\multicolumn{1}{l|}{BiGCN}                   & \multicolumn{1}{c|}{37.91{\small$\pm$2.58}}                     & 43.84{\small$\pm$3.75}      & 51.84{\small$\pm$3.77}      & 17.20{\small$\pm$3.14}      & 27.16{\small$\pm$7.04}      \\
\multicolumn{1}{l|}{GACL}                    & \multicolumn{1}{c|}{54.01{\small$\pm$1.18}}                     & 56.13{\small$\pm$2.06}      & \second\underline{88.14{\small$\pm$1.94}}      & 13.24{\small$\pm$8.88}      & 38.22{\small$\pm$2.97}      \\
\multicolumn{1}{l|}{PSA-S}                   & \multicolumn{1}{c|}{\second\underline{59.36{\small$\pm$1.73}}}                     & \
\best\textbf{92.35{\small$\pm$0.91}}      & 45.81{\small$\pm$4.10}      & 36.23{\small$\pm$4.69}      & \second\underline{52.66{\small$\pm$2.97}}      \\
\multicolumn{1}{l|}{PSA-M}                   & \multicolumn{1}{c|}{58.97{\small$\pm$0.87}}                     & \second\underline{88.30{\small$\pm$0.56}}      & 41.83{\small$\pm$2.62}      & \second\underline{42.14{\small$\pm$2.08}}      & 52.47{\small$\pm$2.03}      \\
\multicolumn{1}{l|}{FADE}                     & \multicolumn{1}{c|}{\best\textbf{71.81{\small$\pm$2.50}}}                     & 56.80{\small$\pm$1.44}      & \best\textbf{92.10{\small$\pm$1.34}}      & \best\textbf{66.42{\small$\pm$2.17}}      & \best\textbf{63.68{\small$\pm$1.97}}      \\ \bottomrule
\end{tabular}}
\caption{Metrics $\pm$ STD (\%) comparison under our experiment setting, averaged over 10 runs. The highest results are highlighted with \colorbox[HTML]{C5DFF8}{\textbf{bold}}, while the second highest results are marked with \colorbox[HTML]{E6FFFD}{\underline{underline}}}
\vspace{-25pt}
\label{tb2}
\end{table}

\begin{table}[]
\renewcommand\arraystretch{0.95}
\resizebox{\linewidth}{!}{
\begin{tabular}{lccccc}
\toprule
\multicolumn{6}{c}{Twitter16}                                                                                            \\ \midrule
\multicolumn{1}{l|}{\multirow{2}{*}{Method}} & \multicolumn{1}{c|}{\multirow{2}{*}{Acc.}} & U     & N     & T     & F     \\ 
\multicolumn{1}{l|}{}                        & \multicolumn{1}{c|}{}                     & F1    & F1    & F1    & F1    \\ \midrule                              
\multicolumn{1}{c|}{BERT}                    & \multicolumn{1}{c|}{41.87{\small$\pm$5.60}}          & 45.00{\small$\pm$3.00}          & 52.00{\small$\pm$5.02}          & 43.00{\small$\pm$3.61}          & 52.00\small{±5.30}          \\
\multicolumn{1}{c|}{BiGCN}                   & \multicolumn{1}{c|}{44.29{\small$\pm$1.34}}          & 46.86{\small$\pm$2.90}          & 44.81{\small$\pm$2.34}          & 53.76{\small$\pm$4.49}          & 25.43{\small$\pm$2.97}          \\
\multicolumn{1}{c|}{GACL}                    & \multicolumn{1}{c|}{\second\underline{71.26{\small$\pm$2.18}}}          & 79.73{\small$\pm$1.76}          & \second\underline{81.83{\small$\pm$0.93}}          & 59.68{\small$\pm$7.36}          & \second\underline{58.11{\small$\pm$2.68}}          \\
\multicolumn{1}{c|}{PSA-S}                   & \multicolumn{1}{c|}{65.43{\small$\pm$0.95}}          & \best\textbf{95.05{\small$\pm$0.80}} & 46.66{\small$\pm$1.64}          & 61.22{\small$\pm$1.49}          & 55.62{\small$\pm$2.35}          \\
\multicolumn{1}{c|}{PSA-M}                   & \multicolumn{1}{c|}{61.47{\small$\pm$1.74}}           & \second\underline{93.91{\small$\pm$0.28}}          & 20.97{\small$\pm$8.51}          & \second\underline{62.21{\small$\pm$1.86}}          & 55.08{\small$\pm$3.93}          \\
\multicolumn{1}{c|}{\textbf{FADE}}            & \multicolumn{1}{c|}{\best\textbf{77.72{\small$\pm$0.48}}} & 83.06{\small$\pm$2.26}          & \best\textbf{83.68{\small$\pm$1.35}} & \best\textbf{74.14{\small$\pm$2.19}} & \best\textbf{63.01{\small$\pm$3.90}} \\ \bottomrule
\end{tabular}}
\caption{Metrics $\pm$ STD (\%) comparison under our experiment setting, averaged over 10 runs.}
\vspace{-25pt}
\label{tb3}
\end{table}

\vspace{-10pt}
\subsection{Result and Discussion}
We evaluated FADE against GACL using standard metrics like Accuracy, Precision, Recall, and F1. Performance results from Twitter15 and Twitter16, shown in Tables \ref{tb2} and \ref{tb3}, confirm our method's effectiveness with a novel data split. On the PHEME dataset, FADE surpassed the second-best method by 12\%, achieving 60\% accuracy.

BERT ranked lowest, while GCN-based models like BiGCN and GACL faltered in our event-separated setup, with BiGCN averaging only 41.76\% accuracy. PSA also underperformed, limited by its reliance on text only.

FADE excelled across all datasets, outstripping leading methods by up to 12\%. Its success stems from: 1) Enhanced sample generation and graph contrastive learning, 2) Bias mitigation through targeted output adjustments, and 3) Utilizing advanced BERT embeddings.

\vspace{-5pt}
\section{Conclusion}
This paper demonstrates that event-separated data splitting better reflects real-world fake news detection on social media. Current methods struggle to detect fake news in unseen events. To overcome this, we introduce FADE, a robust framework designed for dynamic social media environments. FADE employs adaptive augmentation and graph contrastive learning to enhance a target predictor, which is combined with an event-only predictor for debiasing. Our experiments confirm that FADE surpasses existing methods on three datasets. Research involving LLMs is deferred due to funding and hardware constraints.

\vspace{-5pt}
\section{Acknowledgments}
This work is supported by National Natural Science Foundation of China(62141608,62236010,62372454,62206291).

\balance
\printbibliography


\end{document}